\documentclass[acmsmall,screen,manuscript]{acmart}

\AtBeginDocument{%
  }

\setcopyright{acmcopyright}
\copyrightyear{2023}
\acmYear{2023}
\acmDOI{XXXXXXX.XXXXXXX}

\begin{document}

\title{A Scalable Reinforcement Learning-based System Using On-Chain Data for Cryptocurrency Portfolio Management}

\author{Zhenhan Huang}
\email{huang@ftl.iit.tsukuba.ac.jp}
\author{Fumihide Tanaka}
\email{fumihide.tanaka@gmail.com}
\affiliation{%
  \institution{University of Tsukuba}
  \streetaddress{1-1-1 Tennodai}
  \city{Tsukuba}
  \state{Ibaraki}
  \country{Japan}
  \postcode{305-8577}
}

\renewcommand{\shortauthors}{Huang and Tanaka}

\begin{abstract}
  On-chain data (metrics) of blockchain networks, akin to company fundamentals, provide crucial and comprehensive insights into the networks. Despite their informative nature, on-chain data have not been utilized in reinforcement learning (RL)-based systems for cryptocurrency (crypto) portfolio management (PM). An intriguing subject is the extent to which the utilization of on-chain data can enhance an RL-based system's return performance compared to baselines. Therefore, in this study, we propose CryptoRLPM, a novel RL-based system incorporating on-chain data for end-to-end crypto PM. CryptoRLPM consists of five units, spanning from information comprehension to trading order execution. In CryptoRLPM, the on-chain data are tested and specified for each crypto to solve the issue of ineffectiveness of metrics. Moreover, the scalable nature of CryptoRLPM allows changes in the portfolios' cryptos at any time. Backtesting results on three portfolios indicate that CryptoRLPM outperforms all the baselines in terms of accumulated rate of return (ARR), daily rate of return (DRR), and Sortino ratio (SR). Particularly, when compared to Bitcoin, CryptoRLPM enhances the ARR, DRR, and SR by at least 83.14\%, 0.5603\%, and 2.1767 respectively.
\end{abstract}



%
% Keywords. The author(s) should pick words that accurately describe
% the work being presented. Separate the keywords with commas.


\maketitle

\section{Introduction}
Blockchain networks or platforms, each with its native cryptocurrency (cryptos), are numerous today. Analogously, the blockchain can be compared to a company, while cryptocurrency is akin to its publicly traded shares. On-chain data, or on-chain metrics of a blockchain network, are like a company's fundamentals.

Just as fundamentals disclose significant information about a company, on-chain data provide precise, comprehensive records of a blockchain network. Cryptocurrency valuations are influenced by factors including typical on-chain metrics such as circulating supply, exchange flows, and balance on exchanges. Most on-chain data are real-time, sequentially recorded, capturing operational details and metrics of a specific blockchain network and its native cryptocurrency.

Due to the aforementioned nature of on-chain data, people aspire to utilize and incorporate on-chain data into their systems for price prediction and quantitative trading~\cite{jang2017empirical, saad2019toward, jay2020stochastic, 9650873, casella2023predicting}, since the price of crypto can be determined by multiple factors, e.g., hash rate, a typical on-chain metric. Therefore, the incorporation of on-chain data into quantitative trading systems is naturally expected.

However, such utilization of on-chain metrics in an RL-based system for PM has not been implemented so far~\cite{jiang2017deep,10.1007/978-3-030-19823-7_20,10010940}. The extent to which this utilization could help the systems outperform the baselines in terms of return performance is an intriguing question that remains unanswered.

Hence, we propose CryptoRLPM, a novel and scalable end-to-end RL-based system incorporating on-chain data for cryptocurrency PM. CryptoRLPM, a mid-frequency (10-to-30-minute) PM system, consists of five units covering the process from information comprehension to trading order execution. On-chain metrics are tested and specified for each cryptocurrency, overcoming the issue of metric ineffectiveness. Additionally, we introduce the Crypto Module (CM), based on MSPM~\cite{10.1371/journal.pone.0263689}, to ensure scalability and reusability. Each CM reallocates a single-asset portfolio, including a risk-free asset (cash), necessitating the use of $n$ CMs for an $n$-asset portfolio. This setup enables trained CMs to be reusable for the reallocation of any given portfolios. Furthermore, this setup facilitates CryptoRLPM to allow scalable portfolios, with the underlying cryptocurrencies of the portfolios able to be changed at any time as desired. Backtesting with three portfolios constructed for this study, CryptoRLPM demonstrates positive accumulated rate of return (ARR), daily rate of return (DRR), and Sortino ratio (SR), outperforming all the baseline. Specifically, CryptoRLPM shows at least a 83.14\% improvement in ARR, at least a 0.5603\% improvement in DRR, and at least a 2.1767 improvement in SR, compared to the baseline Bitcoin.

To the best of our knowledge, CryptoRLPM is the first RL-based system adopting on-chain metrics comprehensively for cryptocurrency PM. The benchmarking results indicate that CryptoRLPM robustly outperforms the baselines.

\section{Methodology}
\begin{figure}[!h]
    \centering
       \includegraphics[width=0.7\linewidth]{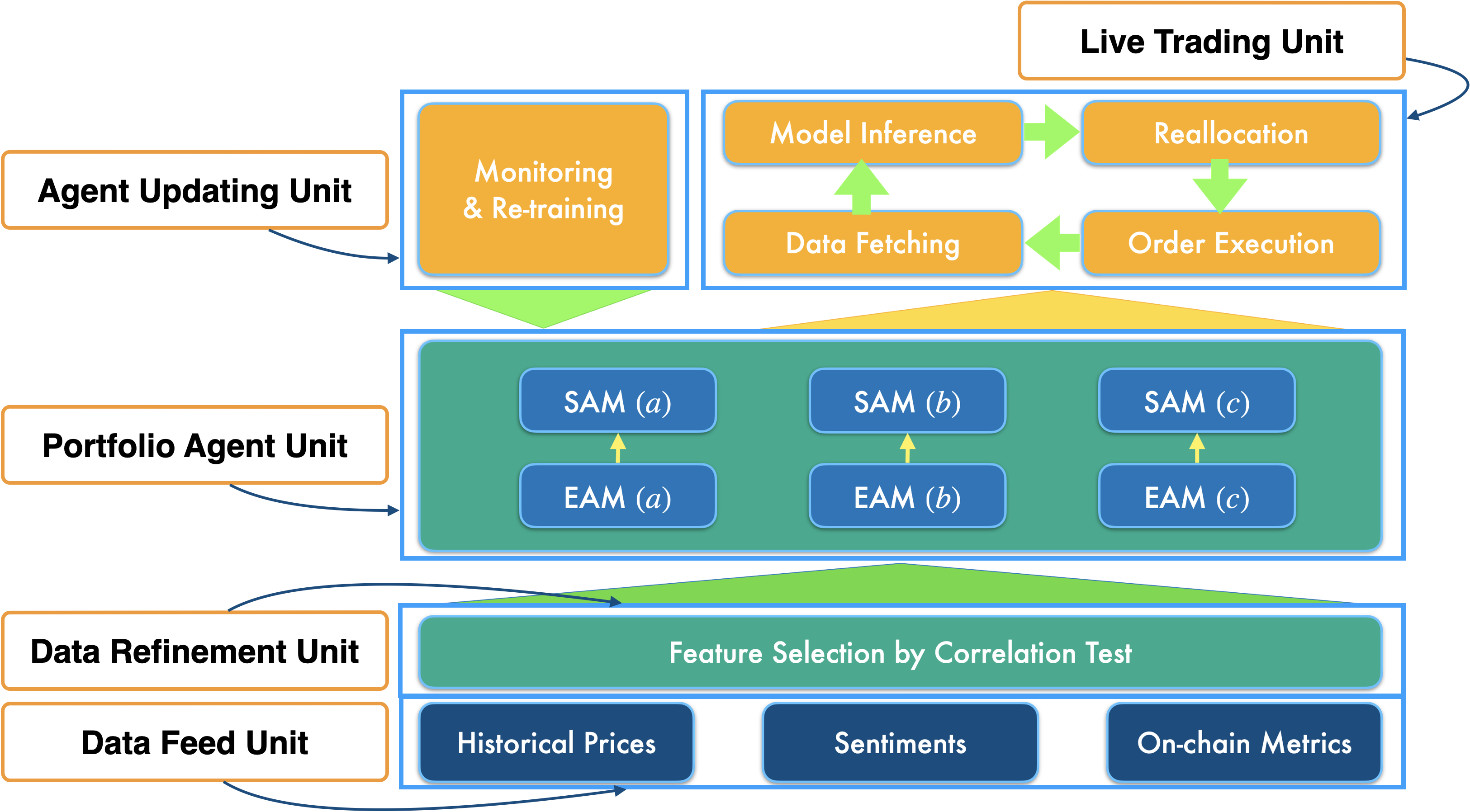}
           \caption{The architecture of CryptoRLPM, illustrating the abstract compositions of each of the five units.}
      \label{fig:10-SD}
\end{figure}

CryptoRLPM is structured into five primary units, which collectively cover the entire process from information comprehension to trading order execution: (i) Data Feed Unit (DFU), (ii) Data Refinement Unit (DRU), (iii) Portfolio Agent Unit (PAU), (iv) Live Trading Unit (LTU), and (v) Agent Updating Unit (AUU). The architecture of CryptoRLPM is illustrated in \autoref{fig:10-SD}.

The five units are interrelated, with each one responsible for at least one distinct task. From a holistic perspective, the Data Feed Unit (DFU) and Data Refinement Unit (DRU) function as the base units related to \textbf{data generation}. The Portfolio Agent Unit (PAU) is responsible for the initial training of RL agents for one or more portfolios. The Live Trading Unit (LTU) and the Agent Updating Unit (AUU) handle the live trading functionality, as well as the maintenance of the agent and the reallocation of portfolios. In the subsequent sections, we will break down and explain the technical details and tasks of each unit. However, the introductions to the LTU and AUU will be rather conceptual, as the purpose of this study is to validate the viability and outperformance of CryptoRLPM. While we do not intend to conduct live trading using CryptoRLPM in this study, we plan to present the implementation of its live trading functionality in future studies.

\subsection{Data Feed Unit (DFU)}
\begin{figure}[!h]
    \centering
       \includegraphics[width=0.45\linewidth]{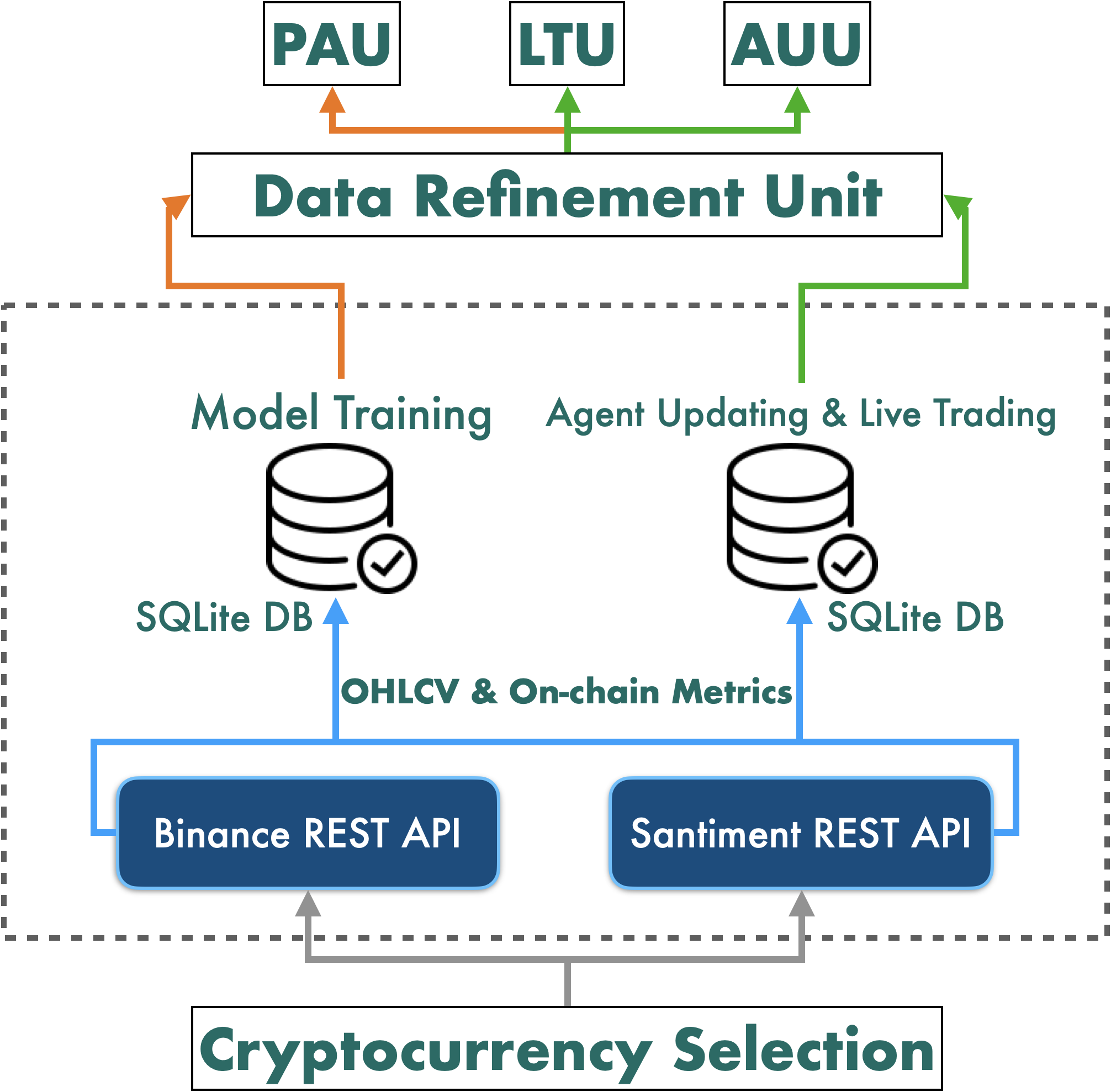}
           \caption{The system design of DFU, with the data flow indicated.}
      \label{fig:10-DFU-SD}
\end{figure}

The Data Feed Unit (DFU) is the most fundamental unit of CryptoRLPM, controlling the acquisition of data both for initial model training and for subsequent ongoing data feed requirements during live trading and model retraining. The system design of DFU is displayed in \autoref{fig:10-DFU-SD}.

\subsubsection{Data Retrieval}
After confirming the portfolio's underlying cryptocurrencies, the DFU retrieves historical price data and on-chain metrics using Binance REST API and Santiment's SanAPI (SanAPI Basic Subscription) respectively~\cite{binance,santiment}. The retrieved data are stored in two separate SQLite databases. The Data Refinement Unit (DRU) will fetch the stored data, and then feed them into the Portfolio Agent Unit (PAU) for model training, Live Trading Unit (LTU) for live trading, and Agent Updating Unit (AUU) for model retraining.

\subsubsection{On-chain Metrics}
On-chain metrics refer to the information generated from the decentralized ledgers of blockchains. For instance, Daily Active Addresses, which represent the number of distinct addresses participating in a transfer of a given crypto on a specific day, indicate the daily level of crowd interaction (or speculation) with that crypto~\cite{daily-active-addresses}. Since most blockchains have their own native cryptocurrencies, the on-chain metrics of a specific blockchain provide insights into its real-time status and ongoing activities.

If we liken a blockchain to a public company, the blockchain's crypto resembles the company's stock, while on-chain metrics mirror its fundamentals. On-chain metrics, due to blockchain's decentralized nature, offer more accurate and transparent measurements than traditional company fundamentals, and are continually public and recorded in real time.

As per the Efficient Market Hypothesis (EMH)~\cite{10.2307/2325486}, a blockchain's crypto valuation presumably reflects all accessible information, including on-chain metrics. Therefore, it is hypothetically to anticipate the incorporation of on-chain data into quantitative trading systems. Nevertheless, to the best of our knowledge, such an integration into an RL-based PM system remains unexplored so far.

\paragraph{Available Metrics:} The on-chain metrics employed in this study are those available under the SanAPI Basic Subscription Plan, and vary depending on different crypto. Given that on-chain and social metrics are often intertwined on API platforms and in practical applications, we do not distinguish between them in this study; both are considered as on-chain metrics.

\subsection{Data Refinement Unit (DRU)}
\begin{figure}[!h]
\centering
\includegraphics[width=0.45\linewidth]{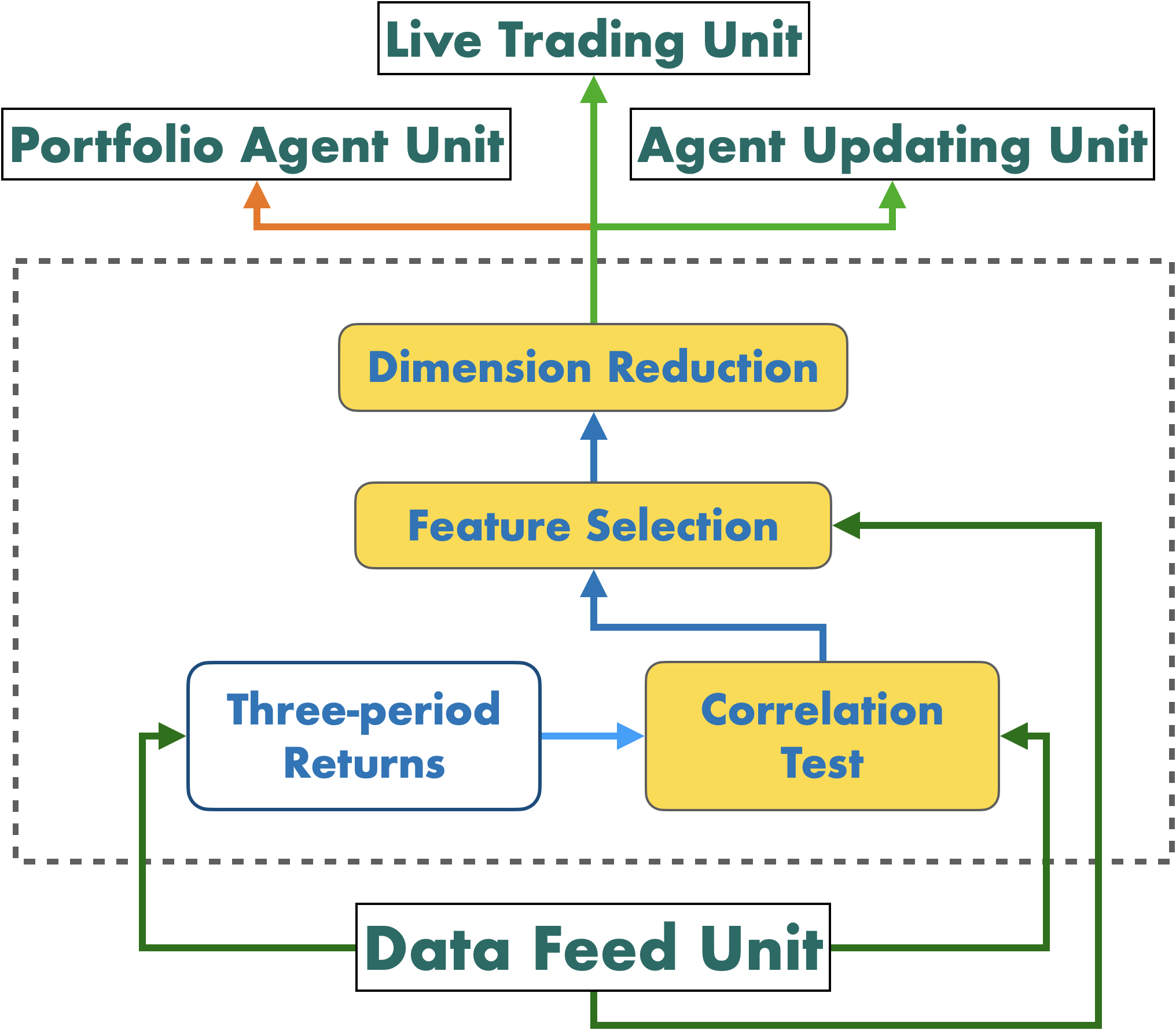}
\caption{The system design of the DRU, illustrating the data flow and component structure.}
\label{fig:10-DRU-SD}
\end{figure}

For any given crypto (e.g., Bitcoin), we conduct correlation tests between the on-chain metrics and three-period returns. \autoref{fig:10-DRU-SD} illustrates the system design of the DRU, as indicated by the dashed line. The term "three-period returns" refers to the percentage change (returns) in a crypto's price over periods of 12, 24, and 48 days. For instance, if we employ Bitcoin's daily OHLCV data, then the three-period returns correspond to the percentage changes in Bitcoin's daily closing prices every 12, 24, and 48 days, respectively.

\subsubsection{Correlation Test for Feature Selection} However, the effective metrics in predicting a particular crypto's price may not be applicable to other cryptos, especially considering that not every crypto has the same set of available metrics. This \textbf{ineffectiveness of metrics} has been barely considered in existing studies.
Thus, in this study, we design a scheme in the DRU to sort the metrics so that they are specified for each crypto in order to mitigate the issue of ineffective metrics.
Our objective is to select \textbf{valid} on-chain metrics from a large pool to construct the environment with which the RL agents interact. To accomplish this, we examine the linear relationship between each of the three-period returns and the on-chain metrics for a specific crypto. This involves determining the Pearson's correlation coefficients between the returns and metrics. The coefficients are divided into three groups, corresponding to the three-period returns. Within each group, the metrics are sorted according to their correlation coefficients, and the highest and lowest five from each group are selected. The selected metrics from all three groups are then ranked by their appearance frequency, and the top-10 metrics are used as valid features to construct the agents' environment in the PAU.

\paragraph{Dimension Reduction:} To further enhance agent learning efficiency, we apply rolling normalization and rolling PCA to the selected metrics for dimension reduction before feeding them into the subsequent units. The principal components that explain at least 80\% of the variance are extracted as the representation of the top-10 metrics and are subsequently fed into the PAU, LTU, and AUU.

\subsection{Portfolio Agent Unit (PAU)}
PAU incorporates the key modules from MSPM\cite{10.1371/journal.pone.0263689}. MSPM is a multi-agent RL-based system designed to address scalability and reusability challenges in RL-based PM. MSPM consists of two key modules, the Evolving Agent Module (EAM) and the Strategic Agent Module (SAM). The EAM leverages a DQN agent to generate asset-specific signal-comprised information. Conversely, the SAM utilizes a PPO agent to optimize the portfolio by connecting with multiple EAMs and reallocating the corresponding assets. As described in~\cite{9905789}, the Strategic Agent Modules (SAMs) of MSPM can be built separately rather than jointly. Namely, each SAM reallocates a single-asset portfolio that includes a risk-free asset (i.e., cash). In a similar vein, within CryptoRLPM, we define a Crypto Module (CM) as a composite module, consisting of an Evolving Agent Module (EAM) and a SAM, dedicated to trading a single crypto. Thus, for example, $n$ CMs will be necessary for the reallocation of an $n$-asset portfolio. With this setup, a trained CM can be integrated into any given portfolio's weighted reallocation alongside other CMs. Furthermore, for efficient training, the EAM within a CM can be optional in certain circumstances, such as when sentiment-included on-chain metrics are fed directly from the DRU to the SAM within the CM. This configuration allows the PAU to be scalable, accommodating variable underlying cryptos in any given portfolio at any time.

\subsubsection{Settings of PAU}
\autoref{fig:10-PAU-SD} illustrates the system design of the PAU, as framed by the dashed line. For the agent training of crypto $x$, on-chain metrics are fed into the DRU from the DFU for selection and dimension reduction. Subsequently, these refined metrics, along with the OHLCV data, are transferred from the DRU to the dedicated EAM of crypto $x$ within the PAU. Alternatively, for the sake of efficient training, the refined metrics can be directly fed into the SAM, as shown by the orange dashed line. In this case, the EAM becomes optional, but the high-quality trading signals from the EAM will not be utilized~\cite{10.1371/journal.pone.0263689}. The signals generated by the EAM, in conjunction with the new OHLCV data, constitute the signal-comprised information that is fed into the SAM of crypto $x$ for decision-making. The trained models are stored separately in the Model Storage. The PAU continues to interact with the AUU for model updates and the LTU for live trading. The EAM and SAM settings are adopted from ~\cite{10.1371/journal.pone.0263689} and ~\cite{9905789}, albeit with modifications. Detailed descriptions and discussions of these modifications follow:

\begin{figure}[!h]
\centering
\includegraphics[width=0.45\linewidth]{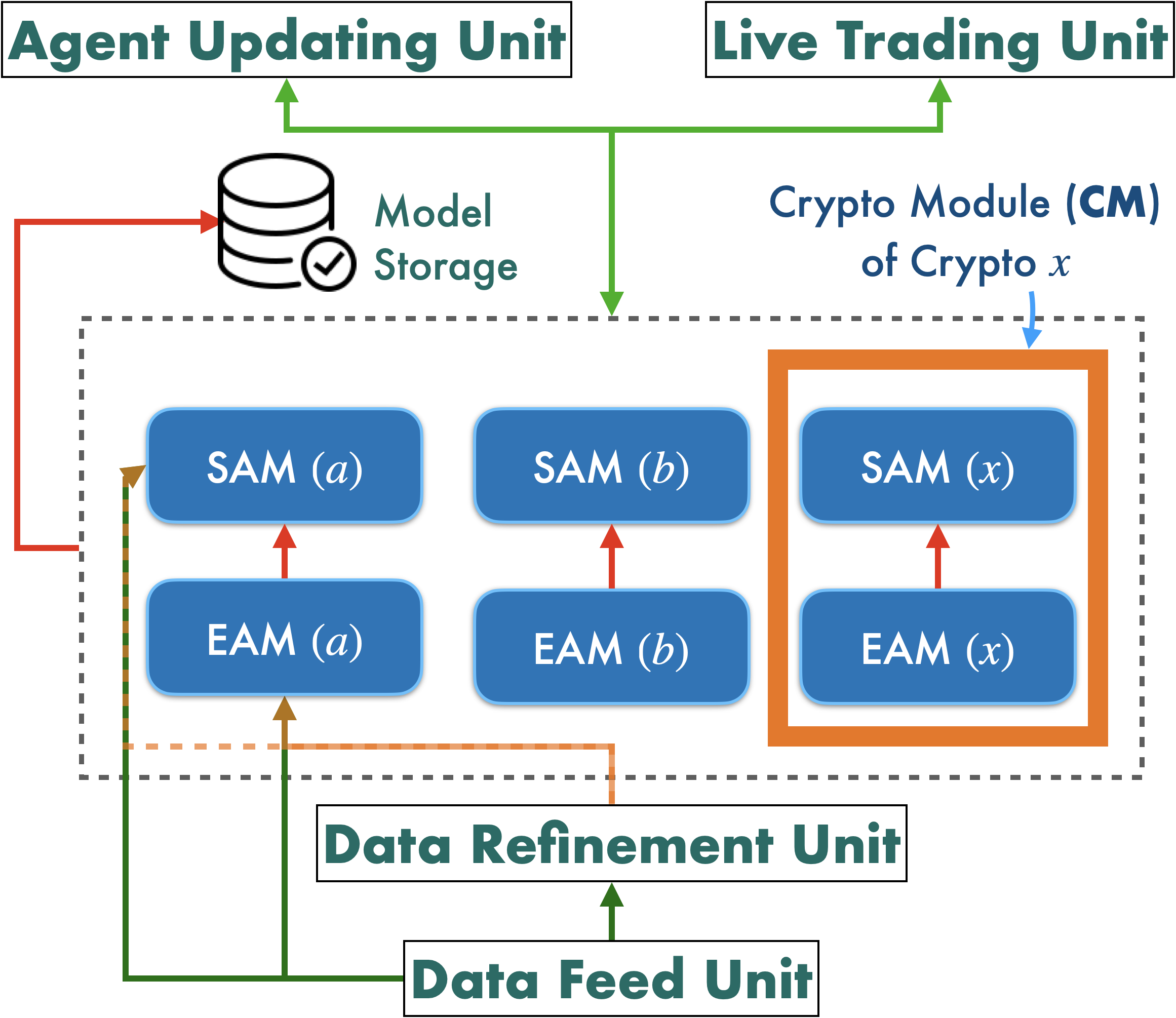}
\caption{The system design of the PAU, detailing the data flow and illustrating the components.}
\label{fig:10-PAU-SD}
\end{figure}

\paragraph{Environment:}
Each crypto-specific CM is composed of a pair: an EAM (optional) and an SAM. The EAM's RL-based agent interacts with an environment formalized by the historical OHLCV and refined on-chain metrics of the designated crypto. The environment for the SAM's RL-based agent is the combination of signals produced by the trained EAM and new OHLCV, or signal-comprised information. Each CM is reusable and periodically retrained by the AUU.

\paragraph{State:}
Within the dedicated CM, the SAM collaborates with the EAM to establish the weight of a specific crypto. The state $v_t$ that the EAM observes at every time step $t$ includes the recent $n$-interval (e.g., 30-minute) OHLCV and refined on-chain metrics of the designated crypto, where $v_t = (s_t, \rho_{t})$, $s_t$ is the $n$-interval OHLCV, and $\rho_t$ represents the refined on-chain metrics from the DRU. In line with the original SAM setting in MSPM, the state $v^+_t$ observed by the SAM at time step $t$ involves the new historical OHLCV stack $s_t$ and the trading signals $a^{sig}t$. Since the SAM in CryptoRLPM is assigned to one crypto, for $v^{+}{t} \in {\mathbb{R}}^{f \times m \times n}$, $f$ denotes the number of features (OHLCV and on-chain metrics), $m = 2$ signifies the designated crypto and cash, and $n$ represents the recent $n$ intervals.

\paragraph{Deep Q-network Agent:}
As introduced previously, both the EAM and SAM use a Deep Q-network (DQN) agent to interact with their environments. Additionally, for the estimates of action-value functions of the EAM and SAM, $Q^{\theta}{EAM}(s_t,a_t)$ and $Q^{\theta}{SAM}(s_t,a_t)$, we continue adopting the settings in~\cite{9905789}, using a 1-D convolutional neural network (CNN) and a simple 4-layer CNN architecture for representation, respectively.

\paragraph{Action Space of EAM:}
At every time step $t$, the DQN agent in the EAM selects an action $a_t$—either {buy, sell, or hold}—for the designated crypto. The actions taken by the EAM establish the crypto's \textbf{\textit{trading signal}}. These actions, stacked with the new OHLCV, are later fed into the SAM within the same CM.

\paragraph{Action Space of SAM:}
In CryptoRLPM, each CM represents a portfolio consisting of the designated crypto and the risk-free asset (cash), which is reallocated by the SAM within it. The SAM of CryptoRLPM assigns full weight to either the risk-free asset or the crypto. In simple terms, at every time step $t$, the action $a_t$ taken by the SAM of CryptoRLPM is a choice from {[0.,~1.] or [1.,~0.]}, indicating the reallocation weight of the portfolio of the designated crypto and cash. With this setting, once an SAM is trained, it can be combined with other CMs and integrated into the voted-weight reallocation of any given multi-crypto portfolio.

\paragraph{Reward Function:}
The reward functions for both the EAM and SAM of CryptoRLPM follow the settings in the original MSPM~\cite{10.1371/journal.pone.0263689,9905789}.

\subsection{Live Trading Unit (LTU)}
\begin{figure}[!h]
    \centering
       \includegraphics[width=0.45\linewidth]{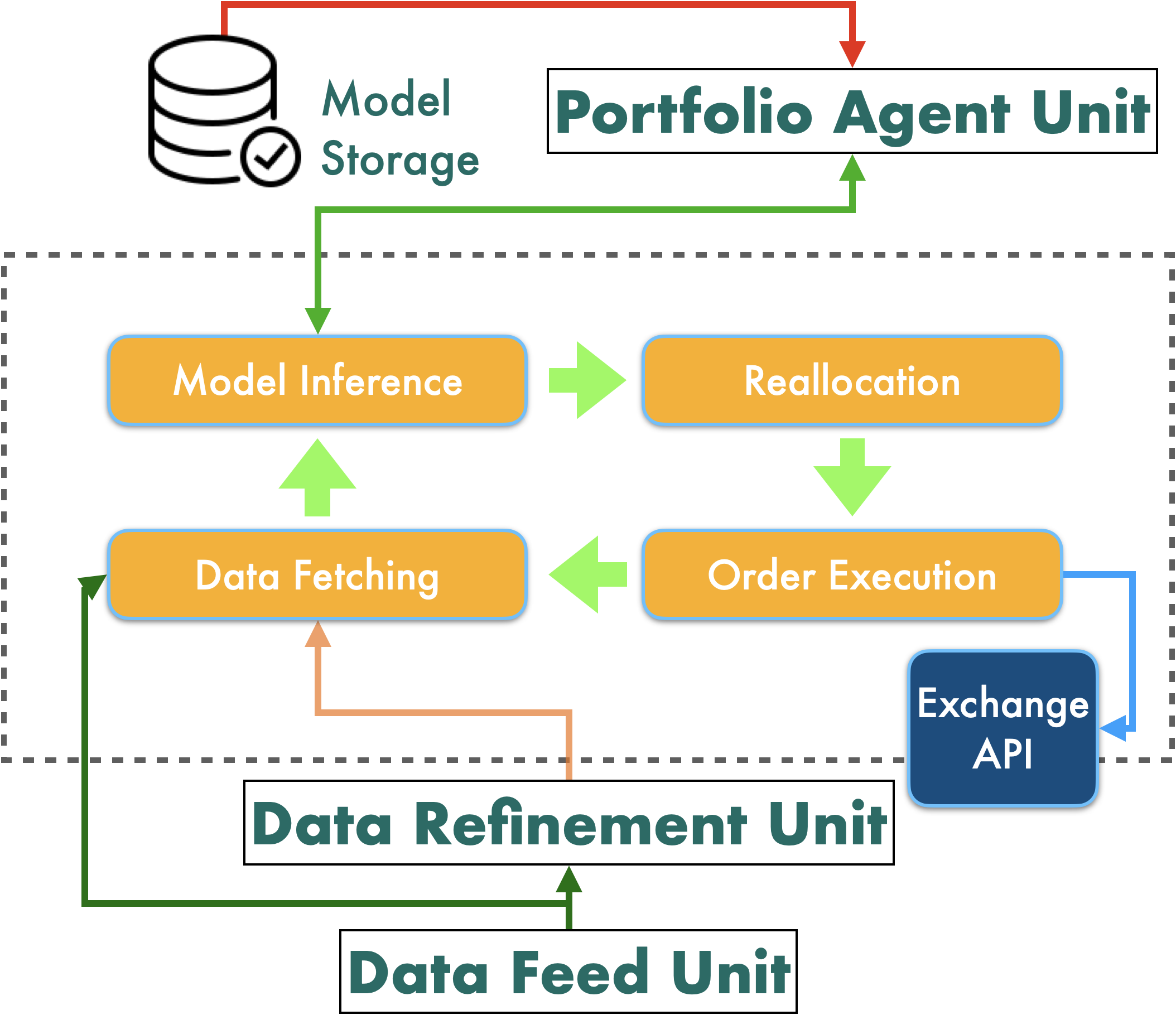}
           \caption{The system design of LTU, with data flow and components illustrated.}
      \label{fig:10-LTU-SD}
\end{figure}

As CryptoRLPM aims to be an end-to-end system for cryptocurrency portfolio management, it naturally incorporates a live trading functionality. In this section, we introduce the Live Trading Unit (LTU) of CryptoRLPM, which manages the live reallocation of the portfolio at 10-to-30-minute intervals. The realization of LTU depends on the APIs of specific exchanges; further implementation details will not be discussed here. The system design of LTU, framed by a dashed line, is shown in \autoref{fig:10-LTU-SD}.

At each $n$ interval, new data are fetched and refined as per the schemes of the first two units. The newly fetched and refined data are fed into PAU for weight inference of the CMs (each corresponding to a designated crypto) in the portfolio. The set $P_t$ comprises the reallocation weights obtained from all $m$ CMs (cryptos) of the portfolio at time step $t$:

\begin{equation}
P_t=\left\{\left(p^{1}_t, \ldots, p^{m}_t\right) \mid p^{i}_t \in {\mathbb{R}}^{2} \text { for every } i \in\{1, \ldots, m\} \right\},
\end{equation}

and the voted weight $w_t$ will be formalized as the reallocation weight of the portfolio at time step $t$

\begin{equation}
w_t = \frac{\sum d^i_t}{m}, \text { for } i \in\{1, \ldots, m\}
\end{equation}

The formalized reallocation weight $w_t$ of the portfolio is transformed into the format required by the designated exchange's API (e.g., Binance). Whenever the portfolio's weight is updated and formatted, a reallocation request is sent to the exchange via their APIs.

\subsection{Agent Updating Unit (AUU)}
The Agent Updating Unit is responsible for scheduled model re-training and unscheduled updates of CMs. After each fixed interval, set in days, the agent models are re-trained, and the portfolio is updated if there are changes to the underlying cryptos, such as scaling or replacing.
% Fig[] shows the system design of AUU.

\section{Experiments}

\subsection{Preliminaries}

\subsubsection{Portfolios}
We propose three portfolios for our experiments:

\begin{enumerate}
    \item Portfolio(a) includes two cryptos:
    \begin{itemize}
        \item Names: Bitcoin (BTC) and Storj (STORJ)
    \end{itemize}
    \item Portfolio(b) includes three cryptos, and shares two cryptos with Portfolio(a):
    \begin{itemize}
        \item Names: Bitcoin (BTC), Storj (STORJ), and Bluzelle (BLZ)
    \end{itemize}
    \item Portfolio(c) includes four cryptos, and shares three cryptos with Portfolio(b):
    \begin{itemize}
        \item Names: Bitcoin (BTC), Storj (STORJ), Bluzelle (BLZ) and Chainlink (LINK)
    \end{itemize}
\end{enumerate}

There are four distinct cryptos denominated by USDT~\cite{tether} (a U.S. dollar equivalent stablecoin) included in the three portfolios. The reusability of CM and scalability of PAU allow the application of the trained crypto-designated CMs to different portfolio-designated PAUs, enhancing efficiency in model training. Consequently, we only need to train four CMs for the four cryptos, and organize these CMs in PAUs to represent and reallocate the three portfolios.

\subsubsection{Data Ranges}
In our experiments, the DFU retrieves historical 6-hour OHLCV data ($s_t$) from~\cite{binance} and on-chain metrics ($\rho_t$) from~\cite{santiment},  which are later refined by the DRU. In this study, the refined metrics are directly fed into the CMs' SAMs from the DRU, leveraging the modularized design of the CM and ensuring efficient training. After that, data is split into three subsets: (i) CM(training) from October 2020 to December 2021; (ii) CM(validation) from January 2022 to February 2022; and (iii) CM(backtesting) from March 2022 to September 2022. Notably, the data ranges for different portfolios vary slightly in practice, based on the varying underlying cryptos. \autoref{tab:10-datasets} lists the ranges of the datasets.

\begin{table}[!ht]
\centering
\caption{Description of Data Ranges}
\begin{tabular}{@{}lrrl@{}}
\hline\noalign{\smallskip}
Purpose & Range \\
\noalign{\smallskip}\hline\noalign{\smallskip}
CM(training) & 2020 Oct$\sim$2021 Dec\\
CM(validation) & 2022 Jan$\sim$2022 Feb\\
CM(backtesting) & 2022 Mar$\sim$2022 Sept\\
\noalign{\smallskip}\hline
\end{tabular}
\label{tab:10-datasets}
\end{table}

\subsubsection{Performance Metrics}
To measure the performance of the CryptoRLPM system and its baselines, we employ three performance metrics: (i) Daily rate of return (DRR), (ii) Accumulated rate of return (ARR), and (iii) Sortino ratio (SR)~\cite{Sortino59}. Higher values for these metrics often indicate higher performance.

\subsection{Results and Discussion}

\subsubsection{Backtesting Performance}

This study primarily aims to validate the feasibility and effectiveness of the proposed system design, and thus, the baselines used for comparison are the historical performances of the underlying cryptos of each portfolio. We conduct backtesting on our CryptoRLPM system and compare its performance against these baselines.

\begin{figure}[!h]
        \centering
           \includegraphics[width=0.6\linewidth]{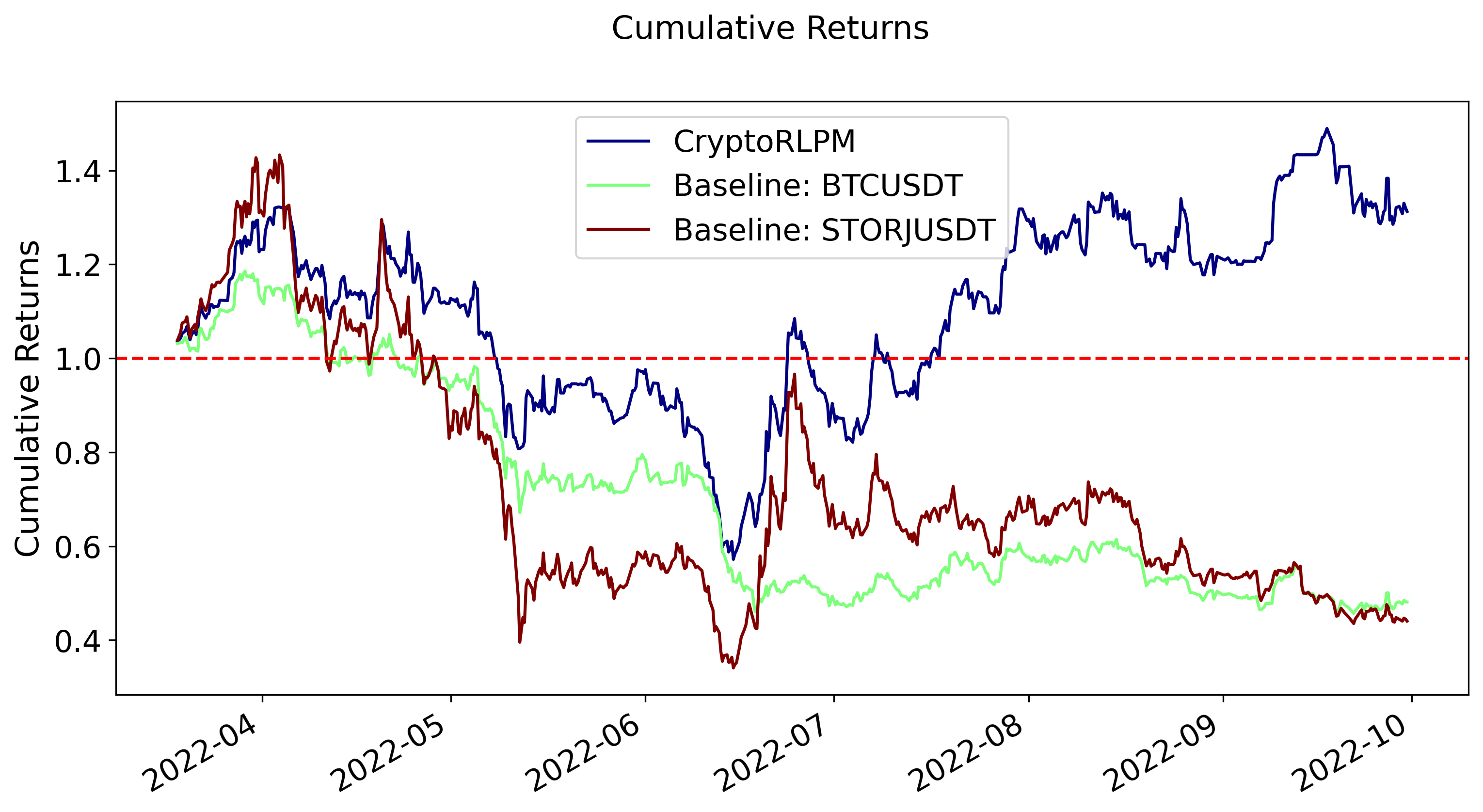}
               \caption{CryptoRLPM outperforms all baselines on Portfolio(a) in terms of the cumulative returns in backtesting.}
          \label{fig:10-BT-Pa}
\end{figure}

\begin{figure}[!h]
        \centering
           \includegraphics[width=0.6\linewidth]{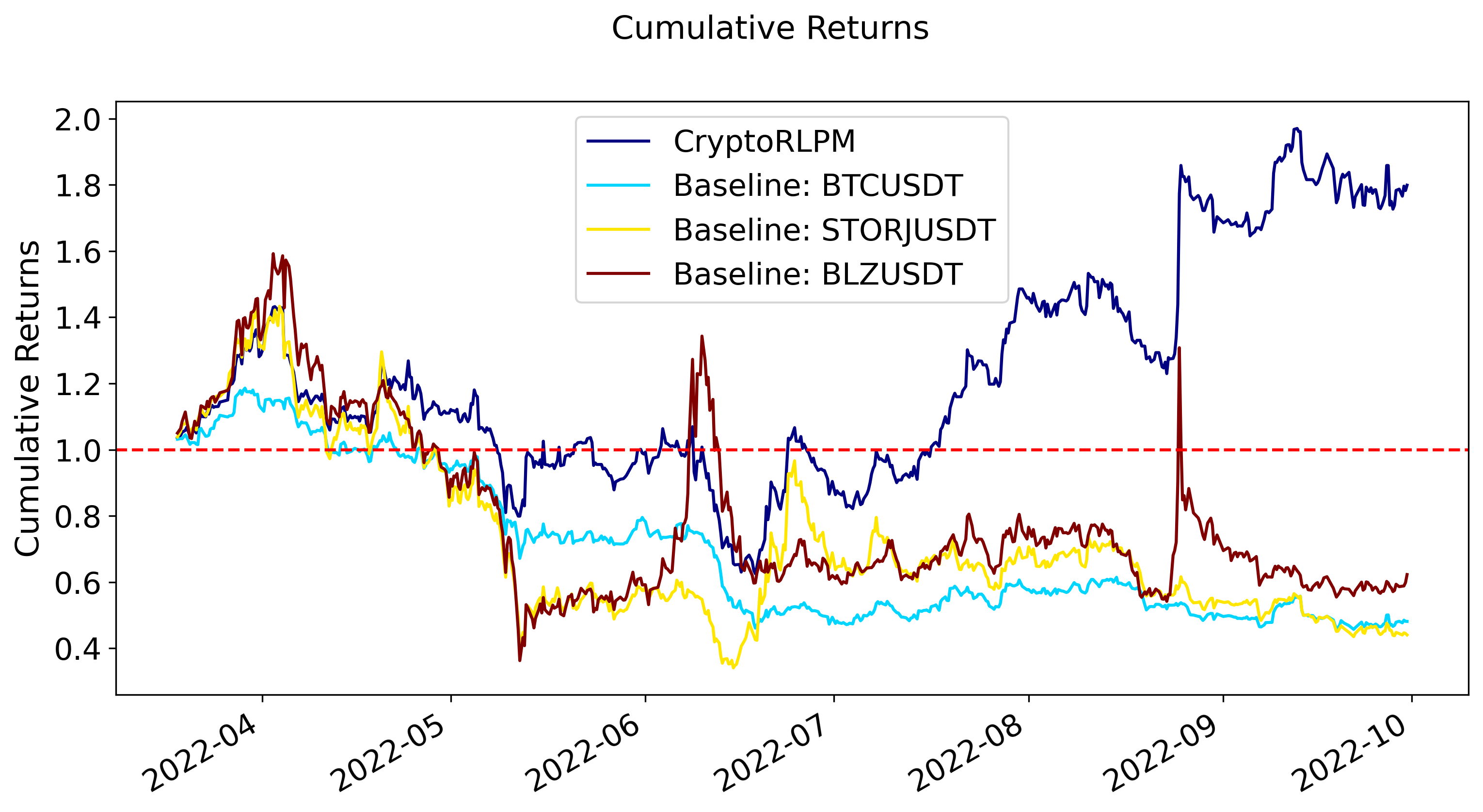}
               \caption{CryptoRLPM outperforms all baselines on Portfolio(b) in terms of the cumulative returns in backtesting.}
          \label{fig:10-BT-Pb}
\end{figure}

\begin{figure}[!h]
        \centering
           \includegraphics[width=0.6\linewidth]{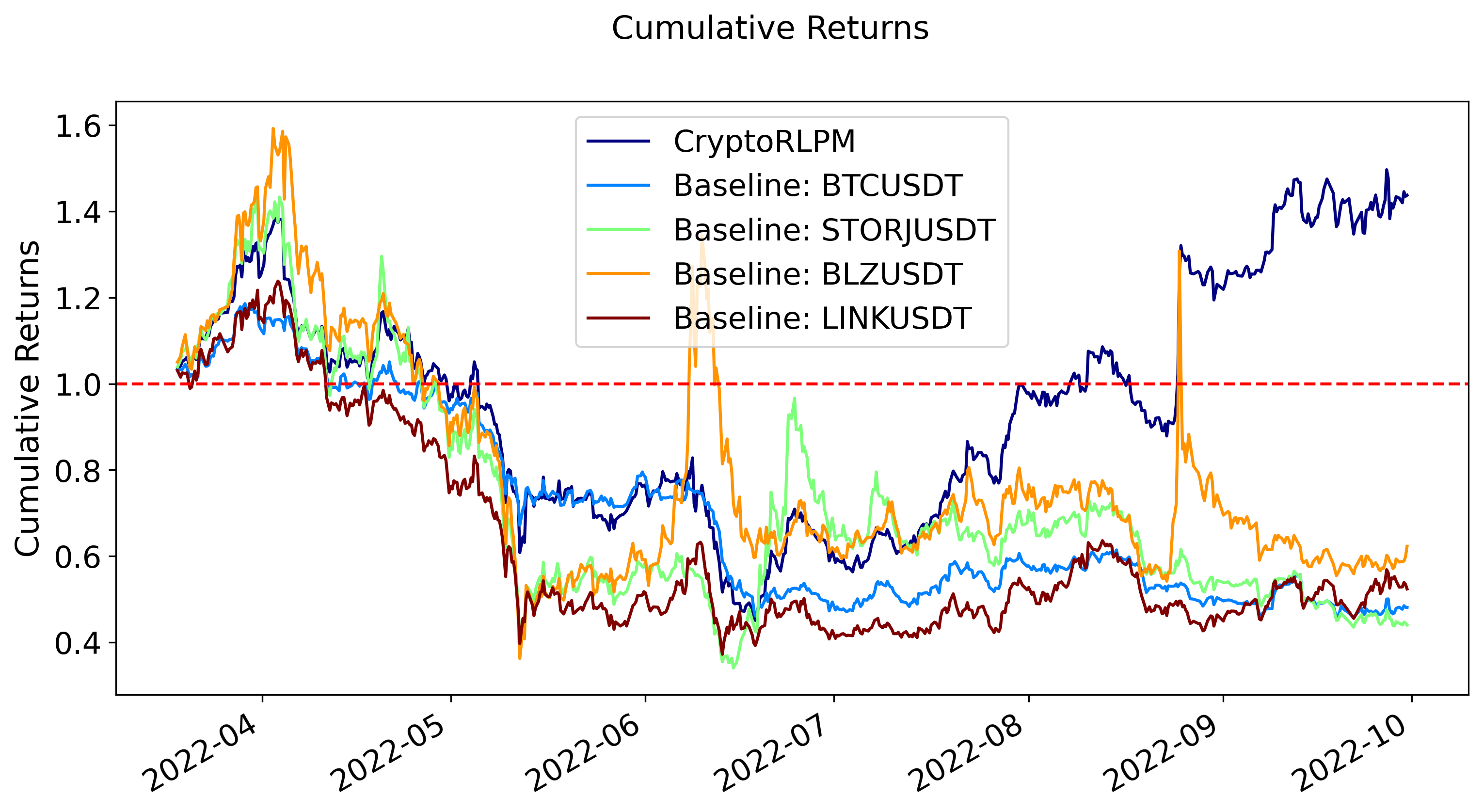}
               \caption{CryptoRLPM outperforms all baselines on Portfolio(c) in terms of the cumulative returns in backtesting.}
          \label{fig:10-BT-Pc}
\end{figure}

As depicted in \autoref{fig:10-BT-Pa}, \autoref{fig:10-BT-Pb}, and \autoref{fig:10-BT-Pc}, CryptoRLPM consistently outperforms the baselines across all three portfolios, achieving positive values in terms of ARR, DRR, and SR, while the baselines yield negative values on ARR. Specifically, when compared to Bitcoin, CryptoRLPM achieves at least a 83.14\% improvement on ARR, a 0.5603\% improvement on DRR, and a 2.1767 improvement on SR. These results demonstrate the strong performance of CryptoRLPM in generating capital returns, and validate the system's potential applicability in crypto PM. \autoref{tab:CryptoRLPM_BT_STAT} details CryptoRLPM's outperformance over the baselines in terms of the ARR, DRR, and SR. \\

\begin{table}[!ht]
\caption{Comparison of backtesting performance of the baselines and CryptoRLPM.}
\resizebox{0.7\linewidth}{!}{%
\begin{tabular}{ccccccc}
\hline
 &  & \textbf{} & \multicolumn{4}{c}{Baselines} \\ \hline
 &  & \textbf{CryptoRLPM} & BTC & STORJ & BLZ & LINK \\ \hline
Portfolio (a) & ARR (\%) & \textbf{31.26} & -51.88 & -55.95 & - & - \\
 & DRR (\%) & \textbf{0.2485} & -0.3118 & -0.1585 & - & - \\
 & SR (\%) & \textbf{0.8709} & -1.3082 & -0.3879 & - & - \\ \hline
Portfolio (b) & ARR (\%) & \textbf{79.87} & -51.88 & -55.95 & -37.72 & - \\
 & DRR (\%) & \textbf{0.4386} & -0.3118 & -0.1585 & 0.1001 & - \\
 & SR (\%) & \textbf{1.3877} & -1.3082 & -0.3879 & 0.2654 & - \\ \hline
Portfolio (c) & ARR (\%) & \textbf{43.71} & -51.88 & -55.95 & -37.72 & -47.60 \\
 & DRR (\%) & \textbf{0.3140} & -0.3118 & -0.1585 & 0.1001 & -0.1785 \\
 & SR (\%) & \textbf{0.9311} & -1.3082 & -0.3879 & 0.2654 & -0.4347 \\ \hline
\end{tabular}}%
\label{tab:CryptoRLPM_BT_STAT}
\end{table}

It is worth noting that CryptoRLPM achieves promising SR for all portfolios, which indicates CryptoRLPM's robust ability at profit-making and adaptability to the ever-changing market.

\subsubsection{Scalability of CryptoRLPM and PAU}
In CryptoRLPM, each crypto is reallocated by a dedicated, decentralized Crypto Module (CM), rendering it a scalable PM system. The scalability means that trained CMs of the underlying cryptos are reusable and changeable for any portfolio.

As an example, for a portfolio $P_{example}$ with three trained CMs/cryptos: $[a,~b,~c]$, to replace crypto $c$ with a new crypto $x$, we train a new CM($x$), unplug the CM($c$), and plug in the trained CM($x$). Scaling up or down a portfolio is even easier. To exclude a crypto, say $b$, we simply unplug CM($b$). To add a new crypto $y$ to $P_{example}$, we plug in a trained CM($y$).

\begin{figure}[!h]
        \centering
           \includegraphics[width=0.6\linewidth]{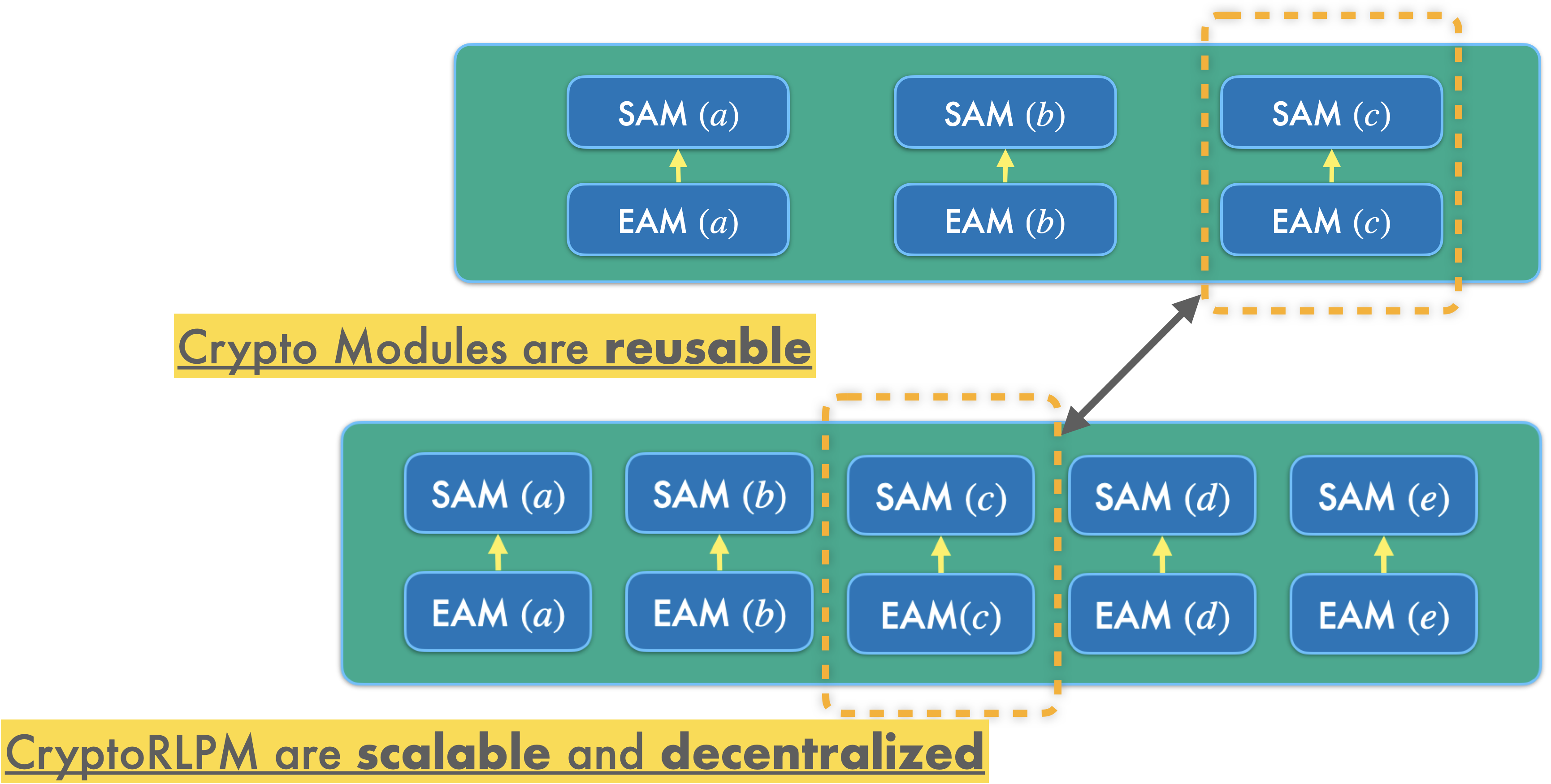}
               \caption{An intuitive illustration featuring the scalability of PAU's architecture. Trained CMs of any cryptos are reusable for different PAUs/portfolios. Trained CMs can be added/plugged to, or removed/unplugged from, any PAUs at will.}
          \label{fig:10-Scalability}
\end{figure}

\autoref{fig:10-Scalability} illustrates the scalability of the architecture. Trained CMs for any cryptos are reusable for different portfolios and can be added or removed at will.

\section{Limitations and Future Work}
In this study, for the efficient training and backtesting of CryptoRLPM, we directly feed refined metrics into PAU from DRU, bypassing EAMs trading signals of CMs. We defer the use of EAMs trading signals to future research, and we anticipate this usage may further enhance CryptoRLPM's performance (ARR, DRR and SR). We also intend to include additional baselines for benchmarking, such as conventional PM strategies, e.g., CRP~\cite{10.1093/rfs/hhm075}, or RL-based methods, e.g. ARL~\cite{liang2018adversarial}. Our focus lies in validating CryptoRLPM's outperformance through backtesting and benchmarking in this study. We plan to present the live trading functionality of CryptoRLPM in future studies.

\section{Conclusion}
We propose CryptoRLPM, a reinforcement learning (RL)-based system incorporating on-chain data for end-to-end cryptocurrency (crypto) portfolio management (PM). CryptoRLPM’s scalability, embodied in its five units, with the reusability of the Crypto Module (CM), enable changes in portfolios' cryptos at any time, demonstrating the system's adaptability to dynamic market conditions. Additionally, we demonstrate CryptoRLPM's ability to efficiently incorporate on-chain metrics for each crypto, overcoming the challenge of metric ineffectiveness. In backtesting with three portfolios, CryptoRLPM consistently delivered positive accumulated rate of return (ARR), daily rate of return (DRR), and Sortino ratio (SR), outperforming all baselines. In comparison to Bitcoin, a prevalent baseline, CryptoRLPM registers at least a 83.14\% improvement in ARR, at least a 0.5603\% enhancement in DRR, and at least a 2.1767 improvement in SR. Our study with its findings highlight the substantial potential of integrating on-chain data into RL-based crypto PM systems to enhance return performance.

\section*{Acknowledgement}
This work was supported by JST SPRING, Grant Number JPMJSP2124.

\bibliographystyle{ACM-Reference-Format}
\bibliography{ref}

\appendix

\end{document}